\documentclass[conference]{IEEEtran}
\ifCLASSINFOpdf
\else
\fi
\usepackage{url}

\usepackage[utf8]{inputenc}
\usepackage{tikz}
\usepackage{listings}
\usetikzlibrary{shapes}
\definecolor{pinegreen}{cmyk}{0.92,0,0.59,0.25}
\definecolor{royalblue}{cmyk}{1,0.50,0,0}
\definecolor{lavander}{cmyk}{0,0.48,0,0}
\definecolor{violet}{cmyk}{0.79,0.88,0,0}
\tikzstyle{cblue}=[circle, draw, thin,fill=cyan!20, scale=0.8]
\tikzstyle{qgre}=[rectangle, draw, thin,fill=green!20, scale=0.8]
\tikzstyle{rpath}=[ultra thick, red, opacity=0.4]
\tikzstyle{legend_isps}=[rectangle, rounded corners, thin,
                       fill=gray!20, text=blue, draw]

\tikzstyle{legend_overlay}=[rectangle, rounded corners, thin,
                           top color= white,bottom color=green!25,
                           minimum width=1.25cm, minimum height=0.6cm,
                           pinegreen]
\tikzstyle{legend_phytop}=[rectangle, rounded corners, thin,
                          top color= white,bottom color=cyan!25,
                          minimum width=1.25cm, minimum height=0.6cm,
                          royalblue]
\tikzstyle{legend_general}=[rectangle, rounded corners, thin,
                          top color= white,bottom color=lavander!25,
                          minimum width=1.25cm, minimum height=0.6cm,
                          violet]

\begin{document}
%
\title{Achlys : Towards a framework for distributed storage and generic computing applications for wireless IoT edge networks with Lasp on GRiSP}

\author{\IEEEauthorblockN{Igor Kopestenski}
\IEEEauthorblockA{ICTEAM Institute\\Université catholique de Louvain\\
igor.kopestenski@uclouvain.be}
\and
\IEEEauthorblockN{Peter Van Roy}
\IEEEauthorblockA{ICTEAM Institute\\Université catholique de Louvain\\
peter.vanroy@uclouvain.be}}


%


\maketitle

\begin{abstract}
Internet of Things (IoT) continues to grow exponentially, in number of devices
and the amount of data they generate.
Processing this data requires an exponential increase in computing power.
For example, aggregation can be done directly at the edge.
However, aggregation is very limited; ideally we would like to do more general
computations at the edge.
In this paper we propose a framework for doing general-purpose edge computing
directly on sensor networks themselves, without requiring external connections
to gateways or cloud.
This is challenging because sensor networks have unreliable communication,
unreliable nodes, and limited (if any) computing power and storage.
How can we implement production-quality components directly on these networks?
We need to bridge the gap between the unreliable, limited infrastructure
and the stringent requirements of the components.
To solve this problem we present Achlys, an edge computing framework
that provides reliable storage, computation, and communication capabilities
directly on wireless networks of IoT sensor nodes.
Using Achlys, the sensor network is able to configure and manage itself directly,
without external connectivity.
Achlys combines the Lasp key/value store and the Partisan communication library.
Lasp provides efficient decentralized storage based on the properties of
CRDTs (Conflict-Free Replicated Data Types).
Partisan provides efficient connectivity and broadcast based on hybrid gossip.
Both Lasp and Partisan are specifically designed to be extremely resilient.
They are able to continue working despite
high node churn, frequent network partitions, and unreliable communication.
Our first implementation of Achlys is on a network of GRiSP embedded system boards.
We choose GRiSP as our first implementation platform because it implements high-level functionality,
namely Erlang, directly on the bare hardware and
because it directly supports Pmod sensors and wireless connectivity.
We give some first results on using Achlys for building edge systems and
we explain how we plan to evolve Achlys in the future.
Achlys is a work in progress that is being done in the context of the LightKone European H2020
research project, and we are in the process of implementing and evaluating a proof-of-concept
application in the area of precision agriculture.

\end{abstract}



%
\IEEEpeerreviewmaketitle

\section{Introduction}

The edge computing paradigm has been widely accepted as an important
concept for sustainability of future Cloud Service Providers (CSPs) and
Mobile Network Operators (MNOs) \cite{belikaidisTrendsChallengesAutonomic2017,
frascolla5GMiEdgeDesignStandardization2017}. It is well acknowledged by both
enterprise and academia as a valid approach and is actively under research
\cite{SmartWorldIoT, kumarComprehensiveStudyIoT2018,
linSurveyInternetThings2017}.
Newer and more performant infrastructures are continuously
elaborated both by CSPs and MNOs \cite{zieglerD6SecondYear}.
Concurrently, IoT devices are getting closer to being actually
\textit{ubiquitous}, i.e., closer to Mark Weiser's idea of \textit{hundreds of
wireless computing devices per person}. This is already true in some scenarios, e.g.,
airplanes generate around 10 TB of data every 30 minutes. Such cases require
very responsive and robust systems for sensor data processing, and could not rely
on remote hosts for it, even if these are close to the edge. Since the Internet of Things is rapidly expanding and devices are becoming more
powerful, IoT applications are putting severe strain on cloud providers.
The edge computing paradigm
is one way to solve this problem by distributing
the workload in a more sustainable way across the whole network.
With this paradigm, computational and storage resources move closer to the edge and IoT
applications are able to preserve their QoS.
Tasks that were previously done in the cloud
are now be delegated to intermediates between datacenters and IoT edge networks.
Existing designs are enabling this by bridging edge networks with intermediate gateways\cite{morabitoLEGIoTLightweightEdge2018}.
However, it is generally considered that the sensor and actuator networks themselves,
such as traditional Wireless Sensor Networks (WSNs), are too
limited and unreliable to do their own management.
Thus even in recent distributed sensor and IoT networks,
a gateway node or a cloud connection is necessary, which adds a single point of failure and increases infrastructure complexity.
If such a point is unable to provide its service, network management becomes impossible
and sensor data cannot be retrieved anymore. And if gateways need to
be permanently available, even short intervals of downtime can disrupt the entire system.
A recent survey has portrayed the full landscape of emerging paradigms
that strive towards the edge and fog principles\cite{yousefpourAllOneNeeds2018}. It describes
a few designs that share common goals with Achlys, such as \textit{mist computing} and suggests that such architectures are a best fit for systems that require 
autonomous behavior, ability for distributed processing directly on IoT devices, little or no Internet connectivity and privacy preservation.

\subsection{The Achlys framework}

This paper presents Achlys, a framework that directly
addresses the problem of general-purpose edge computing.
Achlys increases the
resilience of sensor/actuator edge networks so that they are able to
reliably execute application tasks directly on the edge nodes
themselves. Achlys provides reliable decentralized communication,
storage, and computation abilities, by leveraging CRDTs (Conflict-Free
Replicated Data Types) and hybrid gossip algorithms. This lowers cost,
reduces dependencies, and simplifies maintenance. Our system has no
single point of failure.
Achlys consists of three parts: GRiSP embedded
system boards\footnote{\url{grisp.org}},
a Lasp CRDT-based key/value store, and the Partisan hybrid
gossip-based communication component.
Experimental releases and
source code are available at \textit{\url{ikopest.me}},
and can be deployed on GRiSP boards or run in an Erlang test shell.

Achlys adds a task model to Lasp, which allows applications to be written
by storing both their code and results directly inside Lasp.
In this way, applications are as resilient as Lasp itself.
The task model was first developed as part of
a master's thesis \cite{carlierLaspGrispImplementation2018}.
Application code is replicated automatically by Lasp on all nodes.
From the developer's viewpoint,
Achlys applications are written in a similar way as applications written for
transactional databases.
The developer mindset is that the Lasp database
always contains correct data.
Application tasks can be executed at any time
on any node by the task model.
On every node,
the task model periodically reads tasks from Lasp and executes them.
An executing task
reads data from Lasp, computes the updated data using node-specific sensor
information, actuates node-specific actuators, and stores both the
updated data and an updated task in Lasp.
Because of the convergence properties of CRDTs, the same task can be
executed more than once on different nodes without affecting correctness.
This is a necessary condition for resilience, to keep running despite
node failures and node turnover.

We choose GRiSP because it directly
implements Erlang on the bare hardware, which simplifies system
development, and because it directly supports Pmod sensors and actuators
and has built-in wireless connectivity. Computation and storage abilities
of GRiSP are
limited, but adequate for many management tasks (in fact, with the current
GRiSP infrastructure, our system is
comparable to state-of-the-art laptops released just before the year 2000).
Our current system is a prototype that is able to run applications on
networks of GRiSP boards. With this system we are in the process of
implementing and evaluating a proof-of-concept application of Achlys to
precision agriculture, in collaboration with Gluk Advice BV,
as part of the LightKone European H2020 research project.

\subsection{Motivating example}
We motivate this work with an example related to the precision agriculture
use case we are developing.  We envision a scenario where a farmer
has decided to equip his farms with a Subsurface Drip Irrigation system. Despite that it
is one of the most efficient precision irrigation technologies, it remains
difficult to
have sufficiently precise information about moisture levels, so as
to make optimal use of
the very expensive irrigation infrastructure.
Inadequate irrigation can be extremely
detrimental for production levels, as well as water supplies.

We propose a Self-Sufficient Precision Agriculture Management System
for Irrigation in order to allow the farmer to efficiently irrigate his farms.
In addition to productivity gains, we
intend to offer a solution that is near \textit{zeroconfig}, i.e., it is able
to configure itself with no intervention needed by the farmer.
Finally, we want to provide
a system that manages itself and requires little to no maintenance.
It is possible to connect to the system, but this is only used for setting
policy and is not needed for day-to-day running.
The system's management mechanisms are autonomous,
independent of any third-party providers.

Currently, the farmer's irrigation system
distributes water across the entire zone that is equipped with tubes delivering water to
plantations when the farmer activates a pump. Since moisture levels can vary significantly inside a single farm,
uniform irrigation can be detrimental as some parts will not receive enough
water while in other parts water is wasted and irrigation is above the optimal levels.
Our system is made of a set of distributed edge nodes
that sense moisture levels
and actuate on activation or deactivation of underground irrigation tubes.
The farm is divided into sectors whose moisture levels are
measured by an edge node, and irrigation is adjusted accordingly.
Our solution activates the main irrigation pump and valves when necessary,
and controls the water flow such that once sufficient moisture is achieved,
actuators shut down the water flow of that sector while
irrigation continues in other sectors.
Irrigation decisions are made by an online optimization algorithm
that runs continuously on the edge nodes themselves.
The system thus provides completely autonomous basic management,
without the need for any kind of Internet connection or computer.
The system continuously optimizes its operation
to provide adequate irrigation with minimal cost,
and reconfigures itself whenever it detects a change in configuration.

It is possible to change the irrigation policy by connecting a PC node to the edge network.  This way, we extend the basic autonomous system with
additional features that allow the farmer to use cloud infrastructures
such as storage or high computational power when it is desired.
For example, the edge nodes could be asked to
measure how much water their sector has consumed based on temperatures during that period, and
compute some metrics locally that could be extracted from the edge cluster and stored in the cloud
at the end of each month.
Learning processes applied to that data could again allow farmers to
adjust the system behavior with their computer to gain in efficiency.

%

\subsection{Common challenges}
In contrast to core cloud datacenters, edge networks are composed of large numbers
of heterogeneous devices. Due to the highly dynamic and unpredictable
nature of edge network topologies, nodes can also be temporarily isolated from the network.
For these reasons, implementing desirable features such as reliable computation and storage directly on edge
IoT networks is particularly complex.
Possible solutions are proposed by industry actors such as 5G operators and CSPs, and are generally
based on Points of Presence (POPs) located near client nodes and available through gateways.
In addition, edge applications must be implemented to manage
the limited resources of IoT nodes. Therefore efficient deployments at
the edge require an adequate load balancing mechanism
that ensures that there is no overload on any node.

An important goal of edge computing is to offload efficiently the core of
the network. Since there are several intermediary entities between cloud datacenters and IoT devices
such as servers or smaller datacenters, optimal offloading would be achieved if components of each
layer are able to process some requests autonomously and only rely on higher level nodes when
necessary. Therefore, edge nodes should also strive for maximum independence, and take advantage
of computational and storage resources of IoT devices to complement the edge POP solutions.

Moreover, if edge computing extends the traditional cloud computing paradigm only up to
peripheral POPs, it makes IoT nodes highly dependent on connectivity and exposes single points
of failure. We suggest that the edge paradigm can be implemented in a way that allows offloading
of the core at any level in the global network, even in the most peripheral parts. If IoT
devices are able to provide some basic functionality, then higher level devices can rely on them
to reduce their own workload. And the edge paradigm could therefore maximize the global offloading
since it would distribute the load over all the parts of the edge that are able to perform
tasks such as computations or data storage in a reliable way accordingly to their hardware
resources.

However, despite being standardized to some extent\cite{etsiETSIMEC, etsiETSINGP}, a global production ready
end-to-end solution has not yet been deployed at scales coming close to those of traditional
cloud architectures. There are still many engineering and practical considerations
that must be addressed. In this regard,
the LightKone H2020 European Project aims at providing
a novel approach for general purpose computations at the edge.
LightKone directly addresses the added complexity
due to heterogeneity of IoT devices, which makes a general 
purpose computation model at the edge very attractive.

\subsection{Structure of the article}
The remainder of this article is structured as follows.
Section \ref{contrib} gives a brief overview of
current edge computing state of the art and some key enabling technologies
for Achlys. Section \ref{system} gives
a structural overview of Achlys followed by use case examples
in Section \ref{usecases}.
Finally, Section \ref{conclusion} gives conclusions about the current
state and future evolution of Achlys.

\section{Contributions}
\label{contrib}
In this section, we briefly discuss the contribution of the Achlys application framework in relationship with the global edge computing paradigm.


\subsection{Fault tolerance}
Ensuring fault tolerance is an essential part for generic edge computing\cite{karlssonVegvisirPartitionTolerantBlockchain2018}. In order to fit the vision
of the LightKone project, Achlys strives to guarantee this property. This implies that Achlys must be able to continue functioning even in case of
system failure. These failures can be, but are not limited to :

\begin{itemize}
    \item \textbf{Network partition or intermittent communication} : a node or a set of nodes that are isolated from
the rest of the network must be able to run and to preserve interoperability with
other nodes when the network is repaired.
    \item \textbf{Hardware failure or offline operation} : if a hardware component becomes dysfunctional or goes offline (to save power),
    it should be contained so that the application preserve a maximum amount of features.
\end{itemize}

\subsection{Task model}
Achlys provides a general purpose
task model solution using Erlang higher-order functions. Since Erlang functions are just values (i.e., constants),
they can be copied over a network like any other constant.   Using this ability,
Achlys is able to provide programmers with an API that allows them to easily
disseminate generic tasks in a cluster and be able to retrieve the results if desired.
As handling heterogeneity is a highly complex task for smart services
at the edge\cite{duBigDataPrivacy2018,
jinInformationFrameworkCreating2014,
zanellaInternetThingsSmart2014}, the Achlys prototype can also use this
ability to make the task model homogeneous, despite heterogeneity of the
infrastructure.
This is compatible with a larger vision of future Internet, in which physical
components will be virtualized\cite{espositoCompleteEdgeFunction2017,
ivanovHighlyEfficientRepresentation2017, meiklejohnLoquatFrameworkLargescale2017}

\subsection{Data consistency at the edge}
Conflict resolution is one of the
central problems of distributed and decentralized applications, 
and is the subject of extensive research\cite{baqueroMakingOperationbasedCRDTs, barkerImplementingGarbageCollectedGraph2018, brownBigGerSets2016, almeidaDeltaStateReplicated2018}.
For example, when multiple actors modify the same data entity across a
network partition, what should be done when the partition is repaired?
CRDTs (Conflict-Free Replicated Data Types) provide a solution to this problem.
They are mathematically designed to provide consistent replication with
very weak synchronization between replicas: only eventual replica-to-replica
communication is needed.
The Lasp library uses a wide range of CRDT types for its data storage.
To the developer, Lasp looks like a replicated distributed key/value store
that runs directly on the IoT network.
Section \ref{lasp} gives more information on Lasp and CRDTs.

\section{Overview of the system design}
\label{system}
We now present a more detailed description of the \textit{Achlys} system,
an Erlang\footnote{\url{erlang.org}} implementation of a framework that combines the
power of $\delta$-CRDTs\cite{almeidaDeltaStateReplicated2018} in the
Lasp store\cite{meiklejohnPracticalEvaluationLasp2017, meiklejohnLaspLanguageDistributed2015},
the Partisan communication 
component\cite{meiklejohnPartisanEnablingCloudScale2018},
and the GRiSP Runtime software.
It provides application developers a way to build resilient
distributed edge IoT applications.
We leverage hybrid gossiping and the use of CRDTs in
order to propose a platform that is able to provide reliable services directly
on edge nodes, which are able to function autonomously even when 
no gateway or Internet access is available.

\subsection{GRiSP base}
The GRiSP base board is the embedded system used to deploy Achlys networks in the current
experimental phase. Its main advantage over other hardware is that it has sufficient resources\cite{adamWirelessSmallEmbedded} to run relevant Erlang applications, that is\footnote{
For full specifications please refer to \url{grisp.org}.} :

\begin{itemize}
    \item Microcontroller : Atmel SAM V71, including : \begin{itemize}
        \item ARM Cortex M7 CPU clocked at 300MHz
        \item 64 MBytes of SDRAM
        \item A MicroSD socket
    \end{itemize}
    \item 802.11b/g/n wireless antenna
    \item SPI, GPIO, 1-Wire and UART interfaces
\end{itemize}

\subsection{GRiSP}

\begin{figure}[th]
\centering
\includegraphics[scale=0.17]{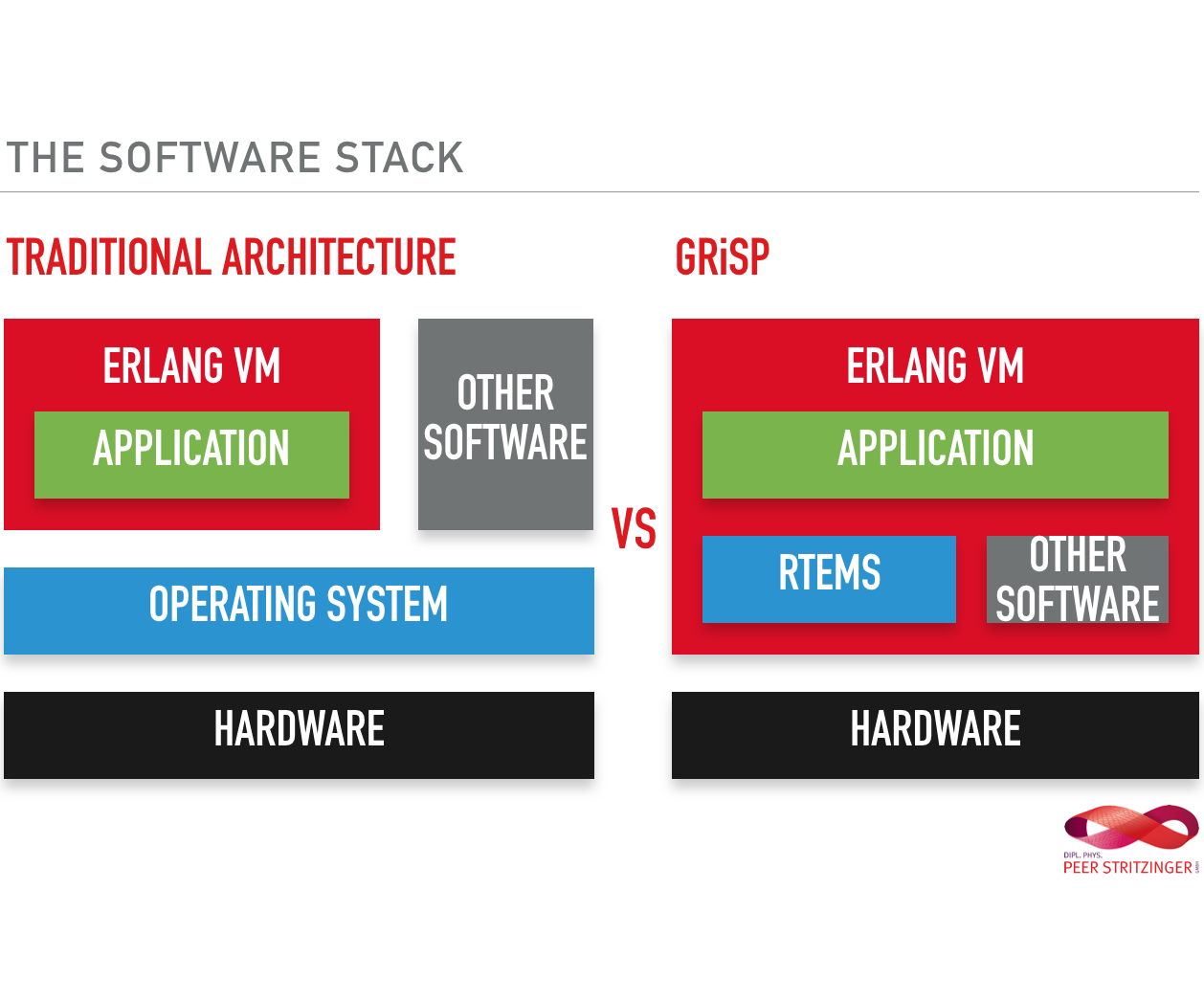}
\caption[GRiSP]{The GRiSP software stack compared to traditional designs. The hardware layer for
GRiSP refers to the \textit{GRiSP base} board that is currently available. Reprinted from GRiSP
presentation by Adam Lindberg.}
\label{fig:grisp}
\end{figure}

Figure \ref{fig:grisp} depicts how the GRiSP architecture is designed. The
RTEMS\footnote{\url{rtems.org}} (RTOS-like set of libraries) component is embedded inside the
Erlang VM and makes it truly run on \textit{bare metal}. Achlys greatly benefits from this unique
design since it allows a much more direct interaction with the GRiSP base hardware. The GRiSP board directly supports Digilent Pmod\footnote{\url{digilentinc.com}} modules. The
latter offer a very wide range of sensing and actuating features that can be accessed at application
level in Erlang.
It is not necessary to write drivers in C in order to add new hardware
features to extend the range of functionalities.

\subsection{Lasp}\label{lasp}
Lasp is a key part of the Achlys framework 
for both storage and computation.
It provides both replicated data and computation,
and guarantees that values will eventually converge on all
nodes\cite{meiklejohnPracticalEvaluationLasp2017, meiklejohnLaspLanguageDistributed2015}.
Since our GRiSP boards run Erlang directly on bare metal, Lasp is a
suitable option for consistency as it runs directly in Erlang\cite{meiklejohnLaspLanguageDistributed2015}.
The Lasp port to GRiSP was initiated in a
master's thesis at UCL\cite{carlierLaspGrispImplementation2018}.
Lasp supports various CRDTs including sets, counters and registers,
all of which do consistent conflict resolution.
For instance, a GCounter is a counter type that can only
be incremented, and when all the operations performed on individual nodes converge, the entire cluster will see the same value,
i.e., the sum of all increments of the counter across all nodes.
This is a very simple example; more complex examples such as sets and
dictionaries are also part of Lasp,
which allow both adding and removing of elements
with the same convergence property.
Achlys is thus able to
handle concurrent modifications and guarantee that all nodes are
eventually consistent, just as on cloud storage services.
This works even on nodes with limited resources and intermittent connectivity.
The only effect of node failures and intermittent connectivity is to slow
down the convergence.
CRDTs in Lasp are implemented using
additional metadata that allows each operation at each node to be taken into consideration.
In fact, the Lasp library uses an efficient implementation of CRDTs
called \textit{delta-based dissemination mode}, which propagates only delta-mutators\cite{almeidaEfficientStatebasedCRDTs2014, almeidaDeltaStateReplicated2018},
i.e., update operations, instead of the full state, to achieve consistency.
This uses significantly less traffic between nodes than a naive implementation
that propagates the full state.

\subsection{Partisan}
Partisan\cite{meiklejohnPartisanEnablingCloudScale2018} is the
communication component used by Lasp to disseminate information
between nodes. It provides a highly resilient
alternative communication layer used instead of the default distributed Erlang communication. This layer
combines the HyParView\cite{leitaoHyParViewMembershipProtocol2007} membership and
Plumtree\cite{leitaoEpidemicBroadcastTrees2007} broadcast algorithms,
to ensure both connectivity and communication, even in extremely
dynamic and unreliable environments. 
Both algorithms use the hybrid gossip approach.
Hybrid gossip is a sweet spot that combines the efficiency of standard
distributed algorithms (e.g., spanning tree broadcast in Plumtree) with
the resilience of gossip. For example, in the case of Plumtree,
the gossip algorithm is used to repair the spanning tree.

Partisan comes with a set of configurable peer service modules that are
each suited for different types of networks. Since the HyParView manager ensures
reliable communication in networks with high attrition rates such as edge clusters,
it is used in our configuration by default. But it can easily be adjusted to match
other types of topologies, and also enables hosts that can be members of multiple
clusters to use the optimal peer service. This makes it suitable for clusters of IoT nodes that are able to communicate with each other despite unreliable networks, but also
to be able to communicate with other cluster types such as star or mesh topologies
where reliable communication does not require the same amount of effort. Partisan is therefore flexible as well as resilient,
and we are still able to configure edge nodes to
communicate with servers, reliable clients or gateways without overhead that would be generated using HyParView in stable networks.
\begin{figure}
    \begin{tikzpicture}[auto, thick, scale=0.85]

        \node[cloud, fill=gray!20, cloud puffs=16, cloud puff arc= 100,
        minimum width=5cm, minimum height=1.75cm, aspect=1] at (0,2) {};

        \foreach \place/\x in {{(-2.5,0.3)/1}, {(-1.75,-0.55)/2},{(-1.2,0.55)/3},
        {(-0.75,-0.7)/4}, {(-0.25,0)/5}, {(0.25,0.7)/6}, {(0.75,-0.3)/7},
        {(1.5,0)/8},{(2.5,0.4)/9}}
        \node[cblue] (a\x) at \place {};

        \path[thin] (a1) edge (a2);
        \path[thin] (a1) edge (a3);
        \path[thin] (a2) edge (a3);
        \path[thin] (a3) edge (a6);
        \path[thin] (a2) edge (a4);
        \path[thin] (a5) edge (a6);
        \path[thin] (a5) edge (a4);
        \path[thin] (a5) edge (a2);
        \path[thin] (a5) edge (a7);
        \path[thin] (a6) edge (a7);
        \path[thin] (a6) edge (a9);
        \path[thin] (a6) edge (a8);
        \path[thin] (a8) edge (a9);
        \path[thin] (a7) edge (a8);

        \draw[very thick, blue, dashed]
        (1.8,1.2).. controls +(left:2.2cm) and +(down:0.5cm) ..(0.7,-1.1);
        \draw[very thick, blue, dashed]
        (-3,1).. controls +(right:2.2cm) and +(down:0.4cm) ..(-1,-1);
        \node[legend_isps] (i1) at (-2.75,-2) { \( temp_{1}(\Delta,m) \) };
        \node[legend_isps] (i2) at (0,-3) { \( p(\Delta,m) \) };
        \node[legend_isps] (i3) at (2,-2) { \( temp_{2}(\Delta,m) \) };
        \draw[-latex, thick, red] (i1) -- (-2.75,0);
        \draw[-latex, thick, red] (i2) -- (0,-0.7);
        \draw[-latex, thick, red] (i3) -- (2,0);

        \foreach \place/\i in {{(-2.5,2.3)/1},{(-1.75,1.45)/2},{(-1.2,2.55)/3},
        {(-0.75,1.3)/4}, {(-0.25,2)/5}, {(0.25,2.7)/6}, {(0.75,1.7)/7},
        {(1.5,2)/8},{(2.5,2.4)/9}}
        \node[qgre] (b\i) at \place {};

        \foreach \i in {1,...,9}
        \path[rpath] (a\i) edge (b\i);

        \node[legend_general] at (0,4){\textsc{Deployed Achlys Cluster}};
        \node[legend_overlay] at (4.5,3){\textsc{Partisan}};
        \node[legend_overlay] at (4.5,2){\textsc{Overlay}};
        \node[legend_phytop] at (4,0.5){\textsc{Physical}};
        \node[legend_phytop] at (4,-0.5){\textsc{Topology}};
    \end{tikzpicture}
    \caption[Network topology]{An example of an Achlys network measuring temperature and pressure.}
    \label{fig:cluster}
\end{figure}
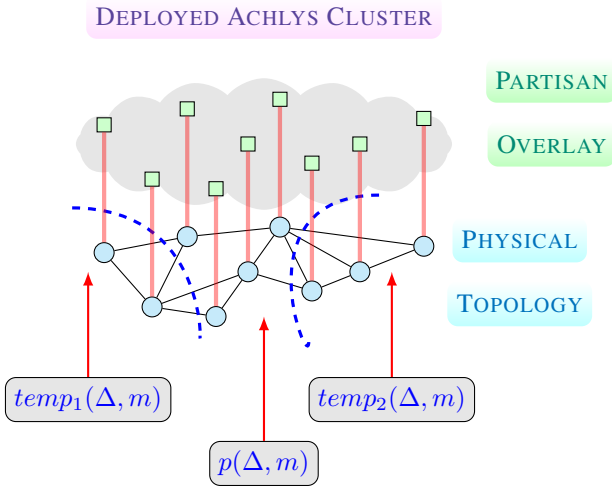

\subsection{Example Achlys network}

Figure \ref{fig:cluster} shows a conceptual overview of a WSN setup of Achlys nodes
that highlights several elements :
\begin{itemize}
    \item The bottom layer consists of functions $temp_{1,2}(\Delta,m)$,$p(\Delta,m)$ that represent
    input streams of data based on environment variables measured by the nodes, in this example
    temperatures and pressure.
    \item The physical topology reflects the real world configuration of the nodes where each edge
    implies that the two vertices are able to establish radio communication.
    \item The virtual overlay that we are able to build using Partisan. Achlys provides
    functions to clusterize GRiSP nodes through Partisan and therefore partitions such as shown
    by dotted blue lines are abstracted away by the eventual consistency and partition tolerance
    properties. Physically isolated parts of the network keep functioning, and seamlessly recover once the links are reestablished.
\end{itemize}

\subsection{Local aggregation}
The vast majority of raw IoT sensor data is usually very short-lived inside systems and ultimately
leads to unnecessary storage. Hence in Achlys we introduce configurable parameters
for aggregation of sensor data. This way programmers can still benefit from distributed storage
but also take advantage of local memory or MicroSD cards to aggregate raw measurements
and propagate mean values. The network loads and global storage volume are thus decreased and
overall scalability is improved.

\subsection{Generic task model}\label{taskmodel}
Achlys provides developers the ability to embed Erlang higher-order functions
through a simple API as shown in Table \ref{tab:api}. This allows building
applications using replicated higher-order
functions. Each node receives tasks and
can locally decide based on load-balancing information and destination targeting
information if it needs to execute it.

We used this model to generate replicated
meteorological sensor data aggregations via generic functions supplied with
specific tasks. A live dashboard of the currently converged view of the data was built and
run on a laptop connected to the GRiSP sensor network.
As long as that web client host was able to reach any node in the network,
it could output its live view of the distributed storage.


\begin{table*}[!t]
  \centering
  \begin{tabular}{|l|l|l|}
    \hline
    \multicolumn{1}{|c|}{Function} & \multicolumn{1}{c|}{Arguments} & \multicolumn{1}{c|}{Description} \\ \hline
    \texttt{add\_task} & \texttt{\{Name, Targets, Fun}\} & \begin{tabular}[c]{@{}l@{}}Adds the task tuple \texttt{\{Name, Targets, Fun\}} to the tasks CRDT\end{tabular} \\ \hline
    \texttt{remove\_task} & \texttt{Name} & \begin{tabular}[c]{@{}l@{}}Removes the task named \texttt{Name} from the tasks CRDT\end{tabular} \\ \hline
    \texttt{start\_task} & \texttt{Name} & \begin{tabular}[c]{@{}l@{}}Starts the task named \texttt{Name}\end{tabular} \\ \hline
    \texttt{find\_and\_start\_task} & \texttt{nil} & \begin{tabular}[c]{@{}l@{}}Fetches any available \\ task from the tasks CRDT and executes it \end{tabular} \\ \hline
    \texttt{start\_all\_tasks} & \texttt{nil} & \begin{tabular}[c]{@{}l@{}}Starts all tasks \\ in the tasks CRDT\end{tabular} \\ \hline
  \end{tabular}
  \vspace{\baselineskip}
  \caption[Generic task model API]{Generic task model API functions.}
  \label{tab:api}
\end{table*}

\section{Additional use cases}
\label{usecases}

\subsection{Live IoT sensor dashboard}
\begin{figure}[th]
\centering
\includegraphics[scale=0.1]{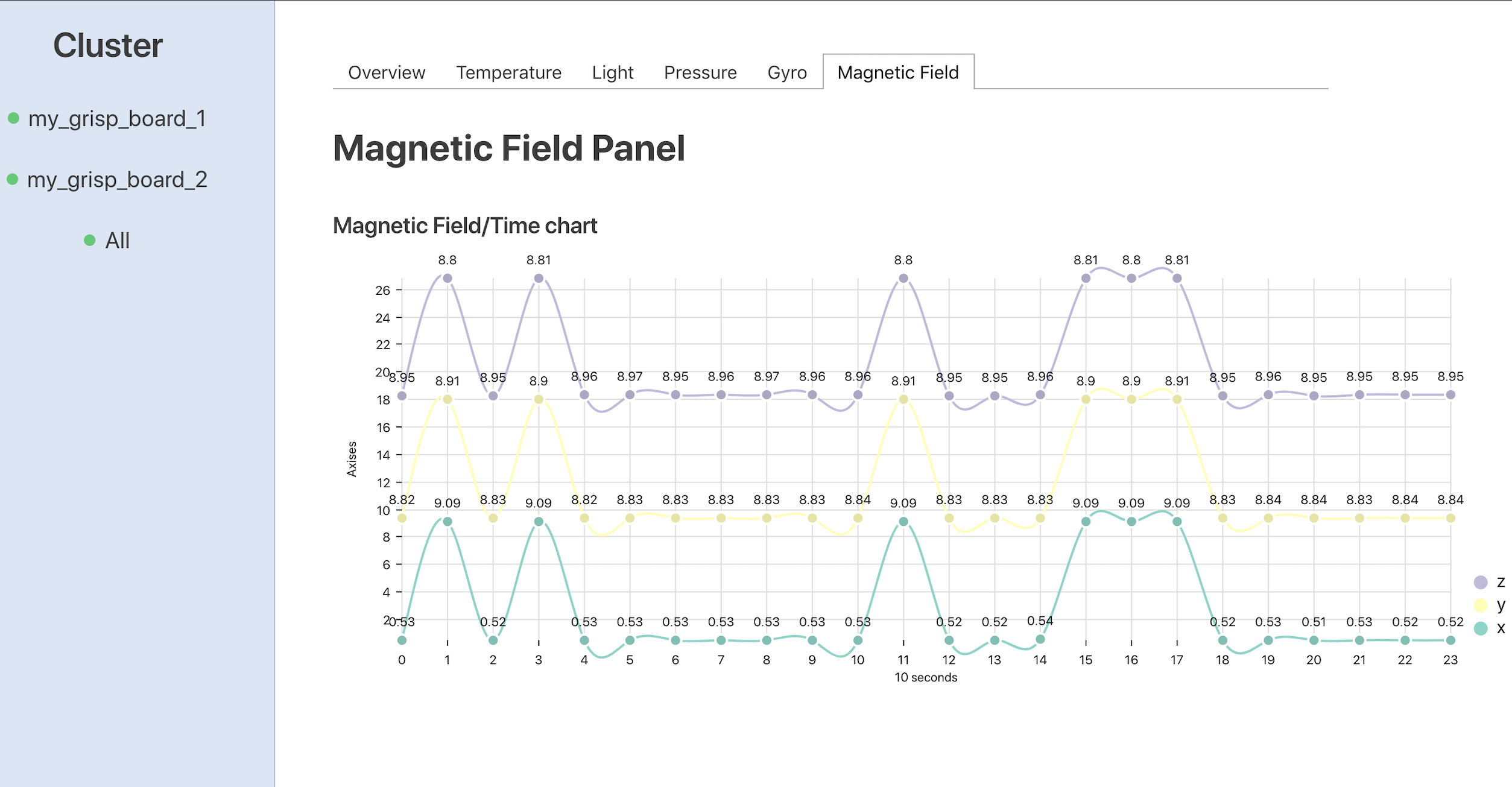}
\caption[Dashboard]{A web client running on the edge and monitoring magnetic field
sensor data from the distributed database.}
\label{fig:dashboard}
\end{figure}

Since our framework is implemented in Erlang, it is also possible to integrate it in
Elixir\footnote{\url{elixir-lang.org}} applications.
Elixir is a programming language that
is built on top of the Erlang runtime system
and adds several very popular web development features. This makes
it possible to implement a web server that runs only at the edge and that can interact with
an entire Achlys cluster as soon as a single node is reachable. Our previous work has already allowed us to display a minimal version of this use case
implementation in the context of the LightKone project. Figure \ref{fig:dashboard} gives a screenshot of a live display of recorded magnetic field data recorded with Digilent Pmod\_NAV modules
attached to GRiSP boards in the cluster. 


Achlys allows the live monitoring to be accessed from any edge node
and it is guaranteed that the distributed database will be consistent regardless of the physical location as long as there is eventually a connection. This use case is a good display of the modular designs that can be implemented with Achlys. We can run isolated applications on IoT networks, add or remove nodes, and easily implement other types of custom nodes that will automatically work the existing cluster. For live monitoring of environmental variables with IoT actuators, the self-configuring and autonomous execution
can be a starting point, and users are free to perform predictive analysis and machine learning y bridging the cluster with the cloud computing services. Once the learning process yields a new set of rules, the users can propagate them to all the network nodes and it will become autonomous again. 


\begin{figure}
\begin{lstlisting}[basicstyle=\ttfamily,language=Erlang]
@ Periodical calls to find
@ available tasks on each node
erlang:send_after(Cycle, self(), trig)

@ Handle the periodical message
handle_info(trigger) ->
    find_and_start_task()
...
@ From single node
@ Create and propagate new task
F = Sense(DeltaInterval, Threshold)
add_task({senseTask, Destinations, F})
\end{lstlisting}
\caption[Generic task model example]{Example usage of the task model API for listening on
the tasks CRDT and running available functions.}
\label{example}
\end{figure}
\subsection{Internet of Battlefield Things}
The U.S. Army Research Laboratory has announced an entire new research
program dedicated uniquely to IoT devices\cite{InternetBattlefieldThings}.
It is focused on deriving the new theoretical models and systems that
will bring the key advantages in military conflicts of the following decades.
The IoBT researchers display a very strong interest in some properties that are
not considered in edge computing IoT architectures designed for commercial use.
The authors describe the IoBT network as dynamic and ubiquitous with a high degree
of pervasiveness, and self-awareness and self-configuration of networks are explicitly
stated as requirements.
Based on the described areas of interest in IoBT, we imagine
the use case of Achlys nodes in forward deployments of ground infrantry squads. These
deployments can lead groups to be isolated in remote areas where no means of communication
with remote operators is possible. During nights, the surroundings must be constantly watched
and soldiers need to spread across areas they are exposed to higher risks once alone.
If soldiers were to be equipped with a set of Achlys nodes, they would be able
to maintain their cluster membership even with all the crew members on the terrain, and
since the sensors can immediately produce data that can indicate distress such as heat or
trembling, the real-time threat analysis can be immediately be propagated to all
the members in the area. In urban combat scenarios, this is particularly desirable as
individuals are often in buildings or confined spaces where they cannot see each other
while Achlys would be able to propagate alarms to all the reachable members. And during all
missions Achlys nodes would record success metrics, environmental variables, and network
topology changes that would be stored during entire operations, and once groups return
to bases they all provide the data to nodes with high computing capacities that can
make predictive analysis that would be used again in future deployments and could
easily be passed between units if they are in communication range. We have observed
that the key features of Achlys that provide reliable communication and storage correspond
to numerous requirements of IoBT since they must remain operational regardless of their
environment.

\section{Conclusion}
\label{conclusion}
We introduce Achlys, a framework for
general purpose edge computing
that runs directly on sensor/actuator networks
with unreliable nodes, intermittent communication,
and limited computation and storage resources.
Our current Achlys prototype is written in Erlang
and runs on a wireless ad hoc network of GRiSP sensor/actuator boards.
Achlys uses the Lasp and Partisan libraries to provide
reliable storage, computation, and communication.
Lasp is a distributed key/value store based on CRDTs,
which is used both for storage (with efficient $\delta$-CRDTs) and
as dynamic management tool to allow dissemination of general-purpose computing functions inside the network.
Partisan is a communication component that
provides highly resilient broadcast and connectivity for dynamic
networks with intermittent connectivity, by using the Plumtree and HyParView
hybrid gossip algorithms.

Because it is written in Erlang,
Achlys can also be used on any infrastructure based on the Erlang runtime.
For example, it can be used on scalable web servers written in Elixir,
because Elixir interoperates seamlessly with Erlang.
Our experiments show encouraging results and validate the feasibility
of our IoT edge computing model. This allows us to focus 
on improving efficiency and usability aspects of Achlys. In particular,
further engineering work will be dedicated to minimize the resource usage of
the framework and its dependencies. We intend to provide a framework that
supports applications on embedded systems in actual deployments, and thus
storage, computation and memory requirements of Achlys need to be carefully
managed. Techniques for compact storage will be investigated such that
we increase the amount of information that is passed through CRDTs while
keeping the size identical.  Furthermore, other optimizations
in terms of networking and self-adaptation will be done in order to
elaborate more intelligent clustering mechanisms. This will be
reported in further work and measurements of efficiency and resilience
with fine-grained adjustments of Partisan's parameters will help
us implement a context-aware networking behavior for Achlys, reducing
unnecessary bandwidth usage and connections. Finally, in order to
reduce application size, we will study if unused modules can
be excluded from the releases deployed on the embedded systems, and if
compression features that are available through Erlang compiler flags
can preserve the features of Achlys while leaving more space for 
applications developed on top.  While we will keep improving Achlys,
it will also be used for proof-of-concept implementations of use case scenarios
in edge computing, and in particular for precision agriculture.


\section*{Acknowledgment}

This work is partially funded by the LightKone European H2020 project
under Grant Agreement 732505.
The authors would like to thank Giorgos Kostopoulos of Gluk Advice BV
for information on precision agriculture.



\bibliographystyle{IEEEtran}
\bibliography{EdgeComputingGroup.bib}
\end{document}